\documentclass[journal]{IEEEtran}

\ifCLASSINFOpdf
\else
\fi

\ifCLASSOPTIONcompsoc
    \usepackage[caption=false, font=normalsize, labelfont=sf, textfont=sf]{subfig}
\else
	\usepackage[caption=false, font=footnotesize]{subfig}
\fi

\usepackage{cite}
\usepackage[version=4]{mhchem}
\usepackage{graphicx}
\usepackage{ifsym}
\usepackage{amsmath,amssymb,amsfonts}
\usepackage[table,xcdraw]{xcolor}

\graphicspath{{figures/}}

\begin{document}

\title{FPGA Synthesis of Ternary Memristor-CMOS Decoders}

\author{Xiaoyuan~Wang,
Zhiru~Wu,
Pengfei~Zhou,
Herbert~H.C.~Iu,
Jason~K.~Eshraghian,
and Sung~Mo~Kang,%
       
\thanks{X.-Y. Wang, Z.~Whu and P. Zhou are with the School of Electronics and Information, Hanzghou Dianzi University, Hangzhou, 310018 China. (e-mail: youyuan-0213@163.com}  
\thanks{H. H. C. Iu is with the School of Electrical and Electronic Engineering, The University of Western Australia, Crawley, WA 6009, Australia.}%
\thanks{J. K. Eshraghian is with the College of Electrical Engineering and Computer Science, University of Michigan, Ann Arbor, Michigan 48109, USA.}%
\thanks{S.M. Kang is with the Department of Electrical and Computer Engineering, University of California, Santa Cruz, Santa Cruz, CA 95064 USA.}}%

\markboth{}%
{}

\maketitle

\begin{abstract}
The search for a compatible application of
memristor-CMOS logic gates has remained elusive, as the 
data density benefits are offset by slow switching speeds and
resistive dissipation. Active
microdisplays typically prioritize pixel density (and therefore
resolution) over that of speed, where the most widely used refresh
rates fall between 25–240 Hz. Therefore, memristor-CMOS logic
is a promising fit for peripheral I/O logic in active matrix displays.
In this paper, we design and implement a ternary 1-3 line decoder
and a ternary 2-9 line decoder which are used to program a
seven segment LED display. SPICE simulations are conducted in
a 50-nm process, and the decoders are synthesized on an Altera
Cyclone IV field-programmable gate array (FPGA) development
board which implements a ternary memristor model designed
in Quartus II. We compare our hardware results to a binarycoded
decimal (BCD)-to-seven segment display decoder, and show
our memristor-CMOS approach reduces the total I/O power
consumption by a factor of approximately 6 times at a maximum
synthesizable frequency of 293.77MHz. Although the speed is
approximately half of the native built-in BCD-to-seven decoder,
the comparatively slow refresh rates of typical microdisplays
indicate this to be a tolerable trade-off, which promotes data
density over speed.


\end{abstract}


\section{Introduction}
\IEEEPARstart{M}{emristor}-CMOS logic has shown much promise with respect to on-chip packing density, and yet, it struggles to be competitive with conventional CMOS processes. There are two common approaches to memristive logic circuits. Stateful logic stores the output signal in-memory, i.e. as the memristor state \cite{kvatinsky2014magic, xie2017scouting, borghetti2010memristive, zheng2019novel, hoffer2020experimental, zhu2013performing, kim2019family},  though it is broadly recognized that stateful logic is burdened with substantial peripheral overhead \cite{hu2018overhead}. The alternative approach typically relies on the nonlinear switching characteristics of a memristor to generate distinguishable output voltage levels that correspond to discrete high and low states \cite{kvatinsky2012mrl, luo2020mtl, papandroulidakis2019practical}. The latter has better integrability with EDA tools as the signal is propagated as a voltage rather than a state, but relies on slow switching processes to generate outputs.

When memristors are used in mixed-signal in-memory computation, device mismatch and data converter overhead appear to be the main bottlenecks \cite{rahimiazghadi2020hardware, eshraghian2019analog, lammie2020memtorch, lin2020adaptive}. The challenges in integrated logic are completely different. The development of a standardized memristor-CMOS logic gate family is largely thwarted by speed and resistive dissipation: the switching speed of memristors is on the order of nanoseconds, whereas the transit delay through a MOSFET channel is on the order of picoseconds in advanced processes. 

Our prior work presented the first experimental demonstration of a complete memristor-CMOS ternary logic family \cite{wang2020high}. 
Data density improvement was between 3.9--25.5 times that of a conventional binary logic CMOS process, depending on the logic gate. This figure of merit accounted for the ability of the memristor-CMOS gates to operate in the ternary-domain, along with integrated chip area where the memristors were fabricated in the back-end-of-the-line (BEOL) of a 50-nm CMOS process.

This benefit came at the expense of significantly slower operation than non-memristive logic gates. While the transit-time in the equivalent CMOS-only process was 9.49ps, there is no memristor logic family that can operate at similar speeds. Although our device made use of fast switching indium-tin-oxide ($\sim$30ns), this is still three orders of magnitude slower than CMOS-only logic. While prior literature often reports picosecond propagation delays, some of these results are idealized simulations that only account for RC delay rather than the slower process of ion transportation \cite{soliman2018memristor}. 

\begin{figure*}[htbp!]
\centering
\includegraphics[scale=0.62]{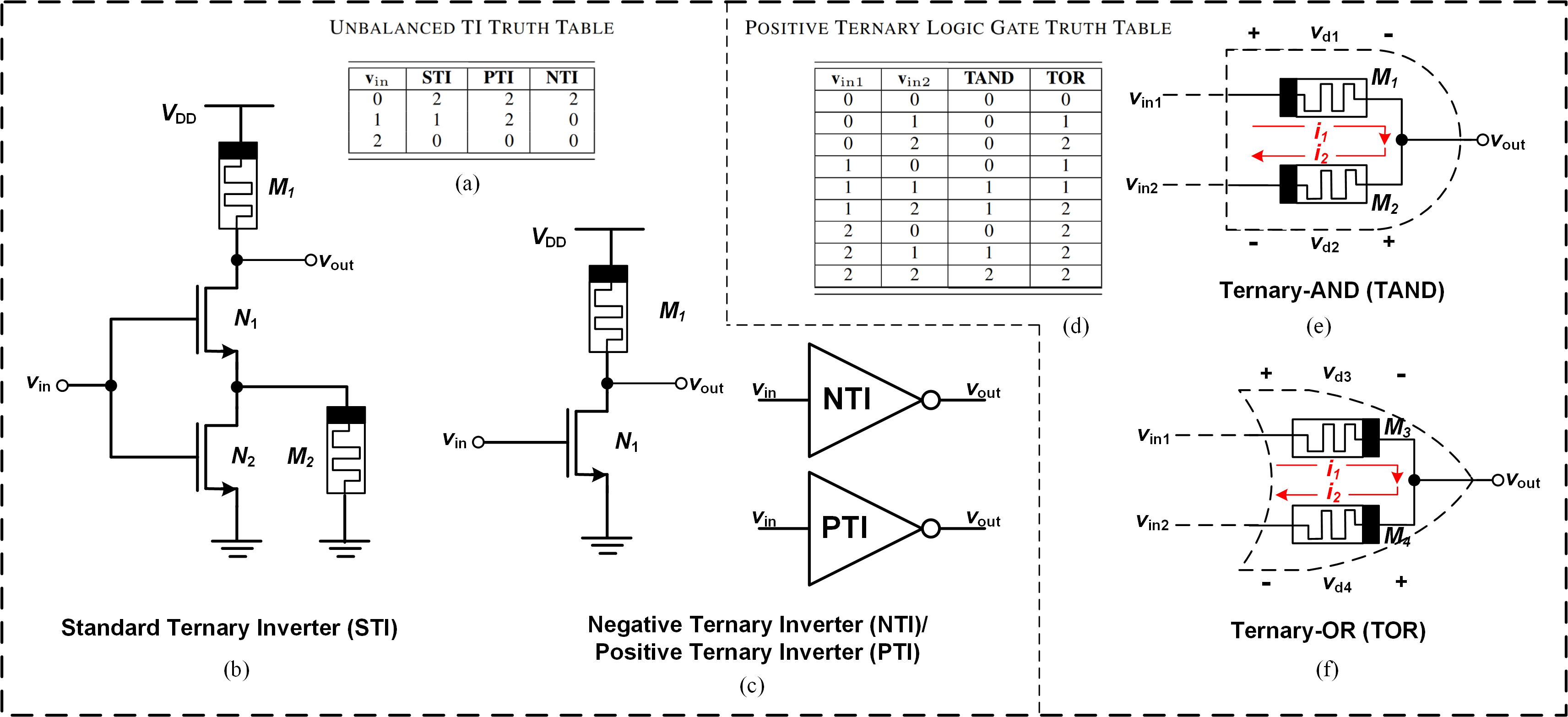}
\caption{Ternary memristor-CMOS logic gates designed and verified in \cite{wang2020high}. (a) Ternary inverter (TI) truth table. (b) Standard ternary inverter (STI) gate. (c) Positive and negative ternary inverter gate (PTI and NTI). The gate level schematics are identical, but the threshold of N$_{\rm 1}$ must be less than V$_{\rm DD}/2$ in the NTI, and above V$_{\rm DD}/2$ for the PTI. This can generally be achieved by appropriate sizing or altering the substrate potential. Exact values are a strong function of the process technology used. (d) Positive ternary logic truth table. Memristor ratioed logic (MRL) gates presented in \cite{kvatinsky2012mrl} are extended from binary to ternary logic. (e) ternary-AND (TAND) gate. (f) ternary-OR (TOR) gate. This is a logically complete family, as all other gates can be implemented using a combination of these primitives.}
\label{fig1}
\end{figure*}

Beyond data encoding, the limited use of multilevel logic can be attributed to the following issues: 1) larger area occupation of multi-level gates, 2) increased propagation delay, and 3) poor noise tolerance. The first of the three issues were alleviated by taking advantage of vertical integration in our prior work \cite{wang2020high}, thus improving area utilization \cite{rahimi2020complementary, eshraghian2018neuromorphic, baek2019memristor}. Potential applications where data density in the form of ternary logic is necessary include various serial links \cite{weber2007multi}, and certain classes of analog-to-digital converters (ADCs) where ternary logic is used in the reduction of quantization errors \cite{guerber201210, kung2018low}. But speed remains critical in serial links and ADCs. 

On the other hand, emerging display applications, including wearable devices, head-mounted virtual and augmented reality displays, require miniaturized, high-density micro-LED arrays. Modern applications of micro-LED arrays are pushing the demand for higher resolution displays, and the continued shrinking of integrated devices has catalyzed their use in biophotonics and optogenetics \cite{mcalinden2019multisite, rabiee2021microarray, mondello2021micro, lin2020development}. In commercial display applications, the refresh rate typically need not exceed 25--240~Hz \cite{menozzi2001crt}. High resolution optoelectronic prostheses can operate effectively at 25~Hz \cite{soltan2018head}. InGaN-based micro-LED arrays have enabled photostimulation of neuron cells \cite{grossman2010multi, jiang2013nitride, ayub2020compact}, where the average firing rate of a neuron may fall between 0.02-20~Hz \cite{hengen2016neuronal}. 

These quantitative values highlight that spatial resolution and pixel density are more critical than speed. Techniques such as flip-chip assembly have been employed to improve micro-LED array density \cite{feng2021p} which brings on the need for higher I/O density, and therefore an exponentially increasing address-range. Memristor-CMOS logic appears to be a natural fit for peripheral I/O logic in active matrix displays, where speed can be sacrificed in lieu of density.

Here, we design and implement a memristor-CMOS ternary 1-3 line decoder and a ternary 2-9 line decoder. We integrate a sequence of combinational ternary memristor-CMOS logic gates using single-stage source-follower buffering to reduce signal attenuation that would otherwise occur from the low output impedance of memristors that are switched on. We conduct SPICE simulations in a 50-nm process, and synthesize the decoders on a Altera Cyclone IV development board using a ternary memristor model designed in Quartus II. We benchmark our hardware results against a BCD-to-seven segment display decoder, and ultimately show our memristor-CMOS approach reduces the total I/O power consumption by a factor of approximately 6 times, with equivalent static power consumption, and a maximum synthesizable frequency of 293.77MHz. 
While the synthesized performance is approximately twice as slow as the BCD-to-seven decoder, we expect the high-data density that comes with memristor-CMOS logic to offset this issue when applied to active matrix microdisplays.

In section 2, we provide a brief background of the ternary logic gates previously presented in \cite{wang2020high} which are scaled up to construct larger combinational memristor-CMOS circuits. We present the design of a memristor-CMOS ternary 1-3 line decoder, followed by the ternary 2-9 line decoder in sections 3 and 4. Section 5 modifies the 2-9 line decoder to drive a seven-segment display decoder. SPICE simulations of each decoder are provided in each section, to validate each design as it is presented. Section 6 provides our experimental FPGA results, followed by a discussion and comparison of power, speed, resource usage between our memristor-CMOS approach and a conventional BCD-to-seven segment decoder.

\section{Memristor-CMOS Ternary Logic Family}

A brief summary of the high-density memristor-CMOS ternary logic family is provided here. The primitive gates are shown in Fig.~1, where unbalanced positive ternary logic is applied in this work: (0, 1, 2) = (GND, V$_{\rm DD}$/2, V$_{\rm DD}$). The line decoders presented in the following sections require buffers between some of the combinational logic stages due to the finite output impedance of the ternary-OR (TOR) and ternary-AND (TAND) gates. A source follower is used to achieve this, where the potential drop from gate to source is recovered in subsequent stages. 

Memristor-CMOS TAND and TOR gates were shown to occupy 6.2\% of silicon area compared to digital CMOS counterparts. The additional data density improvement is quantified by multiplying the improvement by a factor of log(3)/log(2)=1.58 (i.e., three available states in ternary as opposed 2 states in digital). This significant improvement in density motivates their use in display decoders. The switching time was approximately 30~ns, which is not competitive with the transit-time of 9.49~ps at the 50-nm node. The 3-4 orders of magnitude difference in speed means that memristor-CMOS logic is unlikely to be pervasive in modern processor design. But the density advantage can be applied where speed is not paramount. The following sections will use the memristor-CMOS logic primitives from \cite{wang2020high} to build larger combinational decoders that will be used to control a seven-segment LED display, but with the expectation that it can be scaled up further to control a much large active matrix of micro-LEDs.

In general, when the memristor is switched on it forms a pull-up (or pull-down) pathway to the output. If it is switched off, then it will drop the supply (or input) potential. If two memristors form a series path from the supply to ground, then the pair act as a resistive divider regardless of whether they are both on or off, generating an intermediary signal. For brevity, we refer the reader to \cite{wang2020high} for a more detailed description of the operation principles of these gates.

\section{Ternary 1-3 Line Decoder Design}

\begin{figure}[!t]
\centering
\subfloat[]
{
	\includegraphics[scale=0.5]{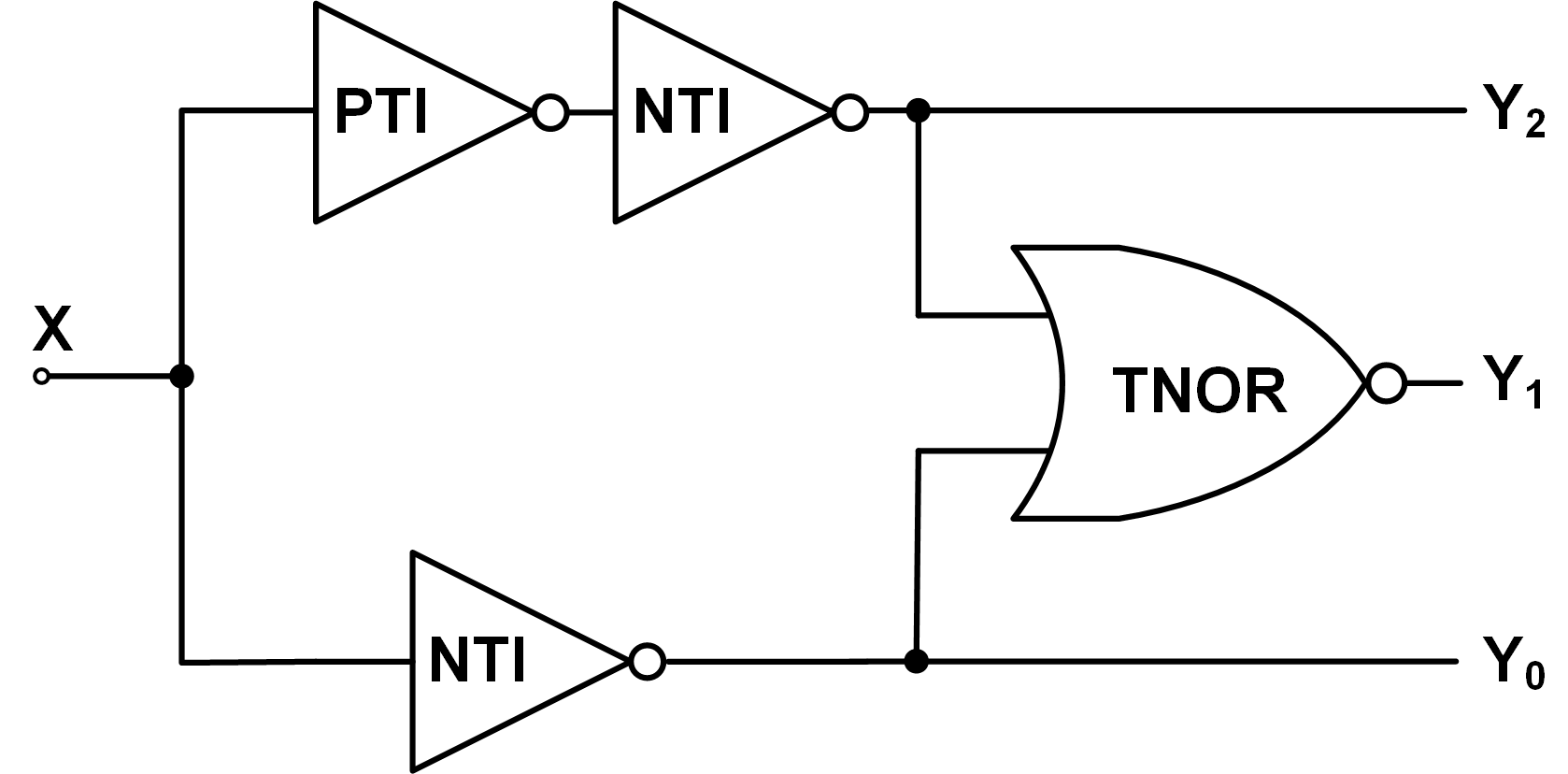}
	\label{fig3a}
}\\
\subfloat[]
{
	\includegraphics[scale=0.5]{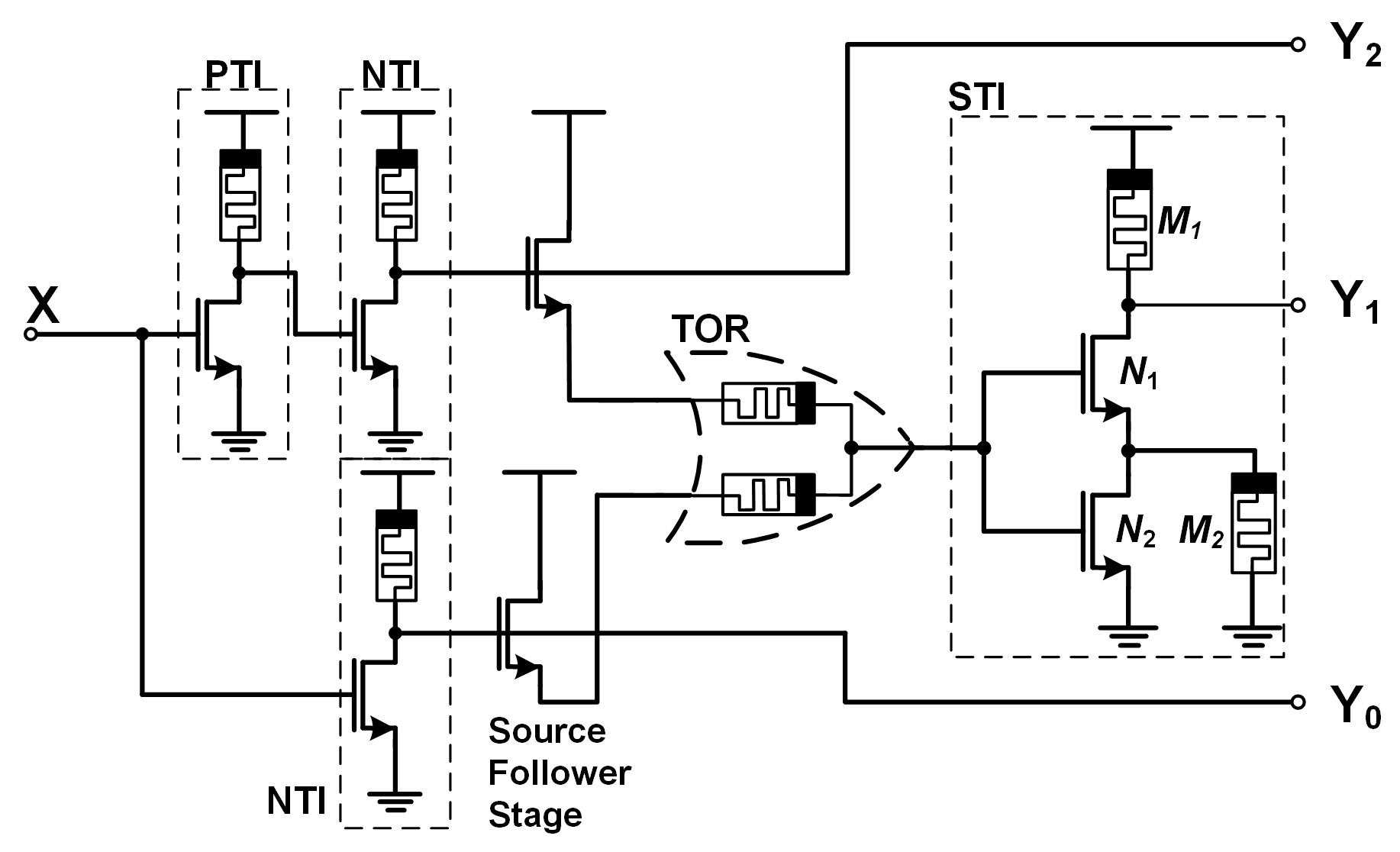}
	\label{fig3b}
}
\caption{Ternary 1-3 line decoder (a) Gate-level schematic. (b) Transistor-level schematic including source-follower buffering to reduce circuit loading that would otherwise occur due to the low impedance seen from the input of the TOR gate.}
\label{fig3}
\end{figure}

\begin{table}[!t]\scriptsize
\centering\caption{1-3 Ternary Line Decoder Truth Table}
\begin{tabular}{
>{\columncolor[HTML]{FFFFFF}}l 
>{\columncolor[HTML]{FFFFFF}}l
>{\columncolor[HTML]{FFFFFF}}l
>{\columncolor[HTML]{FFFFFF}}l}
\hline\vspace{-6pt}
&   \\ \hline
\multicolumn{1}{|c|}{\cellcolor[HTML]{FFFFFF}\textbf{X}} & 
\multicolumn{1}{c|}{\cellcolor[HTML]{FFFFFF}\textbf{Y$_2$}} 
& \multicolumn{1}{c|}{\cellcolor[HTML]{FFFFFF}\textbf{Y$_1$}} 
& \multicolumn{1}{c|}{\cellcolor[HTML]{FFFFFF}\textbf{Y$_0$}} \\ \hline

\multicolumn{1}{|c|}{\cellcolor[HTML]{FFFFFF}0} & \multicolumn{1}{c|}{\cellcolor[HTML]{FFFFFF}0} &
\multicolumn{1}{r|}{\cellcolor[HTML]{FFFFFF}0} &  \multicolumn{1}{r|}{\cellcolor[HTML]{FFFFFF}2}  \\ 
 
\multicolumn{1}{|c|}{\cellcolor[HTML]{FFFFFF}1} & \multicolumn{1}{c|}{\cellcolor[HTML]{FFFFFF}0} &
\multicolumn{1}{r|}{\cellcolor[HTML]{FFFFFF}2} &  \multicolumn{1}{r|}{\cellcolor[HTML]{FFFFFF}0}  \\ 
 
\multicolumn{1}{|c|}{\cellcolor[HTML]{FFFFFF}2} & \multicolumn{1}{c|}{\cellcolor[HTML]{FFFFFF}2} &
\multicolumn{1}{r|}{\cellcolor[HTML]{FFFFFF}0} &  \multicolumn{1}{r|}{\cellcolor[HTML]{FFFFFF}0}  \\ 
 
\hline \vspace{-6pt}
 &  \\ \hline
\end{tabular}%
\end{table}

The function of a ternary decoder is to translate a ternary input code into unary voltage levels of either logic 0 or 2. The truth table of a 1-3 line decoder is shown in Table I, the gate-level schematic is shown in Fig.~2(a), and the transistor-level schematic in Fig.~2(b). It consists of two NTI gates, one PTI gate, and a TNOR gate which consists of a TOR gate in series with a STI gate. A source follower stage interposed before the TOR stage. The PTI and NTI gates are schematically identical, but differ in their voltage thresholds V$_{\rm TH}$. For the NTI gate, V$_{\rm TH} <$V$_{\rm DD}/2$, while for the PTI gate V$_{\rm DD} >$ V$_{\rm TH} > $ V$_{\rm DD}/2$. This can generally be achieved by appropriate sizing or altering the substrate potential, where both are largely dependent on the process technology in use. 

The least significant bit of the output Y$_0$ is generated by a first-stage NTI gate; bit Y$_1$ is taken from the output of the TNOR gate, and the most significant bit Y$_2$ is obtained from the second stage NTI gate. Both Y$_0$ and Y$_2$ are used as the input of the TNOR gate to obtain Y$_1$. SPICE simulations were carried out for the decoder in Fig.~2(b) using Knowm's memristor model \cite{molter2016generalized} using the parameters in Table II. The transistor SPICE models (Level 54 BSIM4) are based on a 50-nm process where V$_{\rm DD}$=1~V. The results are shown in Fig.~3, verifying correct operation of the 1-3 line decoder.

\begin{figure}[t!]
\centering
\includegraphics[scale=0.23]{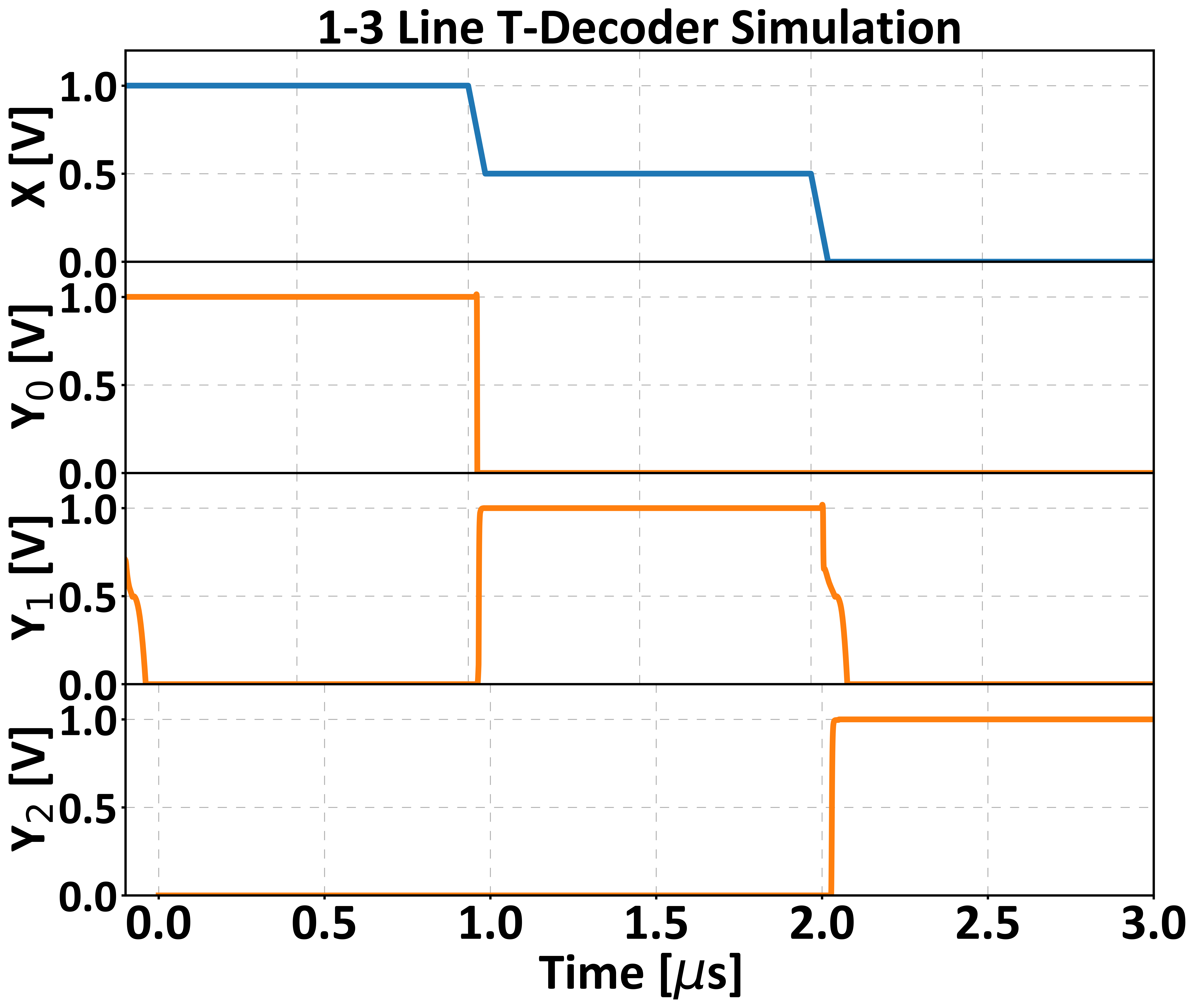}
\caption{SPICE simulation results of the memristor-CMOS ternary 1-3 line decoder from Fig.~2.}
\label{fig3}
\end{figure}

\begin{table}[!t]\scriptsize
\centering\caption{Memristor Model Paramaters}
\begin{tabular}{
>{\columncolor[HTML]{FFFFFF}}l 
>{\columncolor[HTML]{FFFFFF}}l
>{\columncolor[HTML]{FFFFFF}}l}
\hline\vspace{-6pt}
 &   \\ \hline
\multicolumn{1}{|c|}{\cellcolor[HTML]{FFFFFF}\textbf{Parameter}} & 
\multicolumn{1}{c|}{\cellcolor[HTML]{FFFFFF}\textbf{Description}} 
& \multicolumn{1}{c|}{\cellcolor[HTML]{FFFFFF}\textbf{Value}} \\ \hline

\multicolumn{1}{|c|}{\cellcolor[HTML]{FFFFFF}R$_{\rm ON}$, R$_{\rm OFF}$} & 
\multicolumn{1}{c|}{\cellcolor[HTML]{FFFFFF} On/off resistance} & 
\multicolumn{1}{r|}{\cellcolor[HTML]{FFFFFF}500 $\Omega$, 10k $\Omega$}  \\ 
 
\multicolumn{1}{|c|}{\cellcolor[HTML]{FFFFFF}V$_{\rm ON}$, V$_{\rm OFF}$} & 
\multicolumn{1}{c|}{\cellcolor[HTML]{FFFFFF} Set/reset voltage thresholds} & 
\multicolumn{1}{r|}{\cellcolor[HTML]{FFFFFF}0.27 V, 0.27 V}  \\ 

\multicolumn{1}{|c|}{\cellcolor[HTML]{FFFFFF}$\tau$} & 
\multicolumn{1}{c|}{\cellcolor[HTML]{FFFFFF} State variable time constant} & 
\multicolumn{1}{r|}{\cellcolor[HTML]{FFFFFF}500 ps}  \\ 

\multicolumn{1}{|c|}{\cellcolor[HTML]{FFFFFF}T} & 
\multicolumn{1}{c|}{\cellcolor[HTML]{FFFFFF} Temperature} & 
\multicolumn{1}{r|}{\cellcolor[HTML]{FFFFFF}300 K}  \\ 

\multicolumn{1}{|c|}{\cellcolor[HTML]{FFFFFF}$x_0$} & 
\multicolumn{1}{c|}{\cellcolor[HTML]{FFFFFF} State variable initial condition} & 
\multicolumn{1}{r|}{\cellcolor[HTML]{FFFFFF}0}  \\ 
 
\hline \vspace{-6pt}
  &  \\ \hline
\end{tabular}%
\end{table}

\section{Ternary 2-9 Line Decoder Design} 
A ternary 2-9 line decoder can be constructed by routing a pair of ternary 1-3 line decoders (Fig.~2), which is depicted in Fig.~4(a). The corresponding block diagram is in Fig.~4(b). The input signal $A$ passes the most significant ternary bit in, and $B$ passes the least significant ternary bit. The intermediary outputs of the 1-3 decoders are $A_0-A_2$ and $B_0-B_2$, which pass through 2-input TAND gates to generate the nine outputs are are denoted by $Y_0-Y_8$. The truth table of the ternary 2-9 line decoder is shown in Table III, where the following conditions are satisfied: 
\begin{align*}
     Y_8&=A_2B_2 
     &Y_7=A_2B_1 \\
     Y_6&=A_2B_0 
     &Y_5=A_1B_2\\
     Y_4&=A_1B_1 
     &Y_3=A_1B_0\\
     Y_2&=A_0B_2 
     &Y_1=A_0B_1\\
     Y_0&=A_0B_0
\end{align*}

As an example, when the input (2, 1) is applied to ($A$, $B$), the 1-3 line decoders will output ($A_2$, $A_1$, $A_0$)=(2, 0, 0) and ($B_2$, $B_1$, $B_0$)=(0, 2, 0). These intermediate results are passed through the TAND gates to generate (0, 0, 0, 0, 0, 0, 0, 2, 0). 

\begin{table}[!t]\scriptsize
\centering\caption{Ternary 2-9 Line Decoder Truth Table}
\begin{tabular}{
>{\columncolor[HTML]{FFFFFF}}l 
>{\columncolor[HTML]{FFFFFF}}l
>{\columncolor[HTML]{FFFFFF}}l
>{\columncolor[HTML]{FFFFFF}}l
>{\columncolor[HTML]{FFFFFF}}l 
>{\columncolor[HTML]{FFFFFF}}l
>{\columncolor[HTML]{FFFFFF}}l
>{\columncolor[HTML]{FFFFFF}}l
>{\columncolor[HTML]{FFFFFF}}l 
>{\columncolor[HTML]{FFFFFF}}l
>{\columncolor[HTML]{FFFFFF}}l}
\hline\vspace{-6pt}
 &   \\ \hline
 
\multicolumn{1}{|c|}{\cellcolor[HTML]{FFFFFF}\textbf{$A$}} 
& \multicolumn{1}{c|}{\cellcolor[HTML]{FFFFFF}\textbf{$B$}} 
& \multicolumn{1}{c|}{\cellcolor[HTML]{FFFFFF}\textbf{$Y_8$}}
& \multicolumn{1}{c|}{\cellcolor[HTML]{FFFFFF}\textbf{$Y_7$}}
& \multicolumn{1}{c|}{\cellcolor[HTML]{FFFFFF}\textbf{$Y_6$}}
& \multicolumn{1}{c|}{\cellcolor[HTML]{FFFFFF}\textbf{$Y_5$}}
& \multicolumn{1}{c|}{\cellcolor[HTML]{FFFFFF}\textbf{$Y_4$}}
& \multicolumn{1}{c|}{\cellcolor[HTML]{FFFFFF}\textbf{$Y_3$}}
& \multicolumn{1}{c|}{\cellcolor[HTML]{FFFFFF}\textbf{$Y_2$}}
& \multicolumn{1}{c|}{\cellcolor[HTML]{FFFFFF}\textbf{$Y_1$}}
& \multicolumn{1}{c|}{\cellcolor[HTML]{FFFFFF}\textbf{$Y_0$}}\\ \hline

\multicolumn{1}{|c|}{\cellcolor[HTML]{FFFFFF}2} 
& \multicolumn{1}{c|}{\cellcolor[HTML]{FFFFFF}2} 
& \multicolumn{1}{c|}{\cellcolor[HTML]{FFFFFF}2}
& \multicolumn{1}{c|}{\cellcolor[HTML]{FFFFFF}0}
& \multicolumn{1}{c|}{\cellcolor[HTML]{FFFFFF}0}
& \multicolumn{1}{c|}{\cellcolor[HTML]{FFFFFF}0}
& \multicolumn{1}{c|}{\cellcolor[HTML]{FFFFFF}0}
& \multicolumn{1}{c|}{\cellcolor[HTML]{FFFFFF}0}
& \multicolumn{1}{c|}{\cellcolor[HTML]{FFFFFF}0}
& \multicolumn{1}{c|}{\cellcolor[HTML]{FFFFFF}0}
& \multicolumn{1}{c|}{\cellcolor[HTML]{FFFFFF}0}\\ \hline

\multicolumn{1}{|c|}{\cellcolor[HTML]{FFFFFF}2} 
& \multicolumn{1}{c|}{\cellcolor[HTML]{FFFFFF}2} 
& \multicolumn{1}{c|}{\cellcolor[HTML]{FFFFFF}2}
& \multicolumn{1}{c|}{\cellcolor[HTML]{FFFFFF}0}
& \multicolumn{1}{c|}{\cellcolor[HTML]{FFFFFF}0}
& \multicolumn{1}{c|}{\cellcolor[HTML]{FFFFFF}0}
& \multicolumn{1}{c|}{\cellcolor[HTML]{FFFFFF}0}
& \multicolumn{1}{c|}{\cellcolor[HTML]{FFFFFF}0}
& \multicolumn{1}{c|}{\cellcolor[HTML]{FFFFFF}0}
& \multicolumn{1}{c|}{\cellcolor[HTML]{FFFFFF}0}
& \multicolumn{1}{c|}{\cellcolor[HTML]{FFFFFF}0}\\ \hline

\multicolumn{1}{|c|}{\cellcolor[HTML]{FFFFFF}2} 
& \multicolumn{1}{c|}{\cellcolor[HTML]{FFFFFF}1} 
& \multicolumn{1}{c|}{\cellcolor[HTML]{FFFFFF}0}
& \multicolumn{1}{c|}{\cellcolor[HTML]{FFFFFF}2}
& \multicolumn{1}{c|}{\cellcolor[HTML]{FFFFFF}0}
& \multicolumn{1}{c|}{\cellcolor[HTML]{FFFFFF}0}
& \multicolumn{1}{c|}{\cellcolor[HTML]{FFFFFF}0}
& \multicolumn{1}{c|}{\cellcolor[HTML]{FFFFFF}0}
& \multicolumn{1}{c|}{\cellcolor[HTML]{FFFFFF}0}
& \multicolumn{1}{c|}{\cellcolor[HTML]{FFFFFF}0}
& \multicolumn{1}{c|}{\cellcolor[HTML]{FFFFFF}0}\\ \hline

\multicolumn{1}{|c|}{\cellcolor[HTML]{FFFFFF}2} 
& \multicolumn{1}{c|}{\cellcolor[HTML]{FFFFFF}0} 
& \multicolumn{1}{c|}{\cellcolor[HTML]{FFFFFF}0}
& \multicolumn{1}{c|}{\cellcolor[HTML]{FFFFFF}0}
& \multicolumn{1}{c|}{\cellcolor[HTML]{FFFFFF}2}
& \multicolumn{1}{c|}{\cellcolor[HTML]{FFFFFF}0}
& \multicolumn{1}{c|}{\cellcolor[HTML]{FFFFFF}0}
& \multicolumn{1}{c|}{\cellcolor[HTML]{FFFFFF}0}
& \multicolumn{1}{c|}{\cellcolor[HTML]{FFFFFF}0}
& \multicolumn{1}{c|}{\cellcolor[HTML]{FFFFFF}0}
& \multicolumn{1}{c|}{\cellcolor[HTML]{FFFFFF}0}\\ \hline

\multicolumn{1}{|c|}{\cellcolor[HTML]{FFFFFF}1} 
& \multicolumn{1}{c|}{\cellcolor[HTML]{FFFFFF}2} 
& \multicolumn{1}{c|}{\cellcolor[HTML]{FFFFFF}0}
& \multicolumn{1}{c|}{\cellcolor[HTML]{FFFFFF}0}
& \multicolumn{1}{c|}{\cellcolor[HTML]{FFFFFF}0}
& \multicolumn{1}{c|}{\cellcolor[HTML]{FFFFFF}2}
& \multicolumn{1}{c|}{\cellcolor[HTML]{FFFFFF}0}
& \multicolumn{1}{c|}{\cellcolor[HTML]{FFFFFF}0}
& \multicolumn{1}{c|}{\cellcolor[HTML]{FFFFFF}0}
& \multicolumn{1}{c|}{\cellcolor[HTML]{FFFFFF}0}
& \multicolumn{1}{c|}{\cellcolor[HTML]{FFFFFF}0}\\ \hline

\multicolumn{1}{|c|}{\cellcolor[HTML]{FFFFFF}1} 
& \multicolumn{1}{c|}{\cellcolor[HTML]{FFFFFF}1} 
& \multicolumn{1}{c|}{\cellcolor[HTML]{FFFFFF}0}
& \multicolumn{1}{c|}{\cellcolor[HTML]{FFFFFF}0}
& \multicolumn{1}{c|}{\cellcolor[HTML]{FFFFFF}0}
& \multicolumn{1}{c|}{\cellcolor[HTML]{FFFFFF}0}
& \multicolumn{1}{c|}{\cellcolor[HTML]{FFFFFF}2}
& \multicolumn{1}{c|}{\cellcolor[HTML]{FFFFFF}0}
& \multicolumn{1}{c|}{\cellcolor[HTML]{FFFFFF}0}
& \multicolumn{1}{c|}{\cellcolor[HTML]{FFFFFF}0}
& \multicolumn{1}{c|}{\cellcolor[HTML]{FFFFFF}0}\\ \hline

\multicolumn{1}{|c|}{\cellcolor[HTML]{FFFFFF}0} 
& \multicolumn{1}{c|}{\cellcolor[HTML]{FFFFFF}2} 
& \multicolumn{1}{c|}{\cellcolor[HTML]{FFFFFF}0}
& \multicolumn{1}{c|}{\cellcolor[HTML]{FFFFFF}0}
& \multicolumn{1}{c|}{\cellcolor[HTML]{FFFFFF}0}
& \multicolumn{1}{c|}{\cellcolor[HTML]{FFFFFF}0}
& \multicolumn{1}{c|}{\cellcolor[HTML]{FFFFFF}0}
& \multicolumn{1}{c|}{\cellcolor[HTML]{FFFFFF}0}
& \multicolumn{1}{c|}{\cellcolor[HTML]{FFFFFF}2}
& \multicolumn{1}{c|}{\cellcolor[HTML]{FFFFFF}0}
& \multicolumn{1}{c|}{\cellcolor[HTML]{FFFFFF}0}\\ \hline

\multicolumn{1}{|c|}{\cellcolor[HTML]{FFFFFF}0} 
& \multicolumn{1}{c|}{\cellcolor[HTML]{FFFFFF}1} 
& \multicolumn{1}{c|}{\cellcolor[HTML]{FFFFFF}0}
& \multicolumn{1}{c|}{\cellcolor[HTML]{FFFFFF}0}
& \multicolumn{1}{c|}{\cellcolor[HTML]{FFFFFF}0}
& \multicolumn{1}{c|}{\cellcolor[HTML]{FFFFFF}0}
& \multicolumn{1}{c|}{\cellcolor[HTML]{FFFFFF}2}
& \multicolumn{1}{c|}{\cellcolor[HTML]{FFFFFF}0}
& \multicolumn{1}{c|}{\cellcolor[HTML]{FFFFFF}0}
& \multicolumn{1}{c|}{\cellcolor[HTML]{FFFFFF}2}
& \multicolumn{1}{c|}{\cellcolor[HTML]{FFFFFF}0}\\ \hline

\multicolumn{1}{|c|}{\cellcolor[HTML]{FFFFFF}0} 
& \multicolumn{1}{c|}{\cellcolor[HTML]{FFFFFF}0} 
& \multicolumn{1}{c|}{\cellcolor[HTML]{FFFFFF}0}
& \multicolumn{1}{c|}{\cellcolor[HTML]{FFFFFF}0}
& \multicolumn{1}{c|}{\cellcolor[HTML]{FFFFFF}0}
& \multicolumn{1}{c|}{\cellcolor[HTML]{FFFFFF}0}
& \multicolumn{1}{c|}{\cellcolor[HTML]{FFFFFF}2}
& \multicolumn{1}{c|}{\cellcolor[HTML]{FFFFFF}0}
& \multicolumn{1}{c|}{\cellcolor[HTML]{FFFFFF}0}
& \multicolumn{1}{c|}{\cellcolor[HTML]{FFFFFF}0}
& \multicolumn{1}{c|}{\cellcolor[HTML]{FFFFFF}2}\\ \hline

\hline \vspace{-6pt}
  &  \\ \hline
\end{tabular}%
\end{table}

\begin{figure*}[htbp!]
\centering
\includegraphics[scale=0.5]{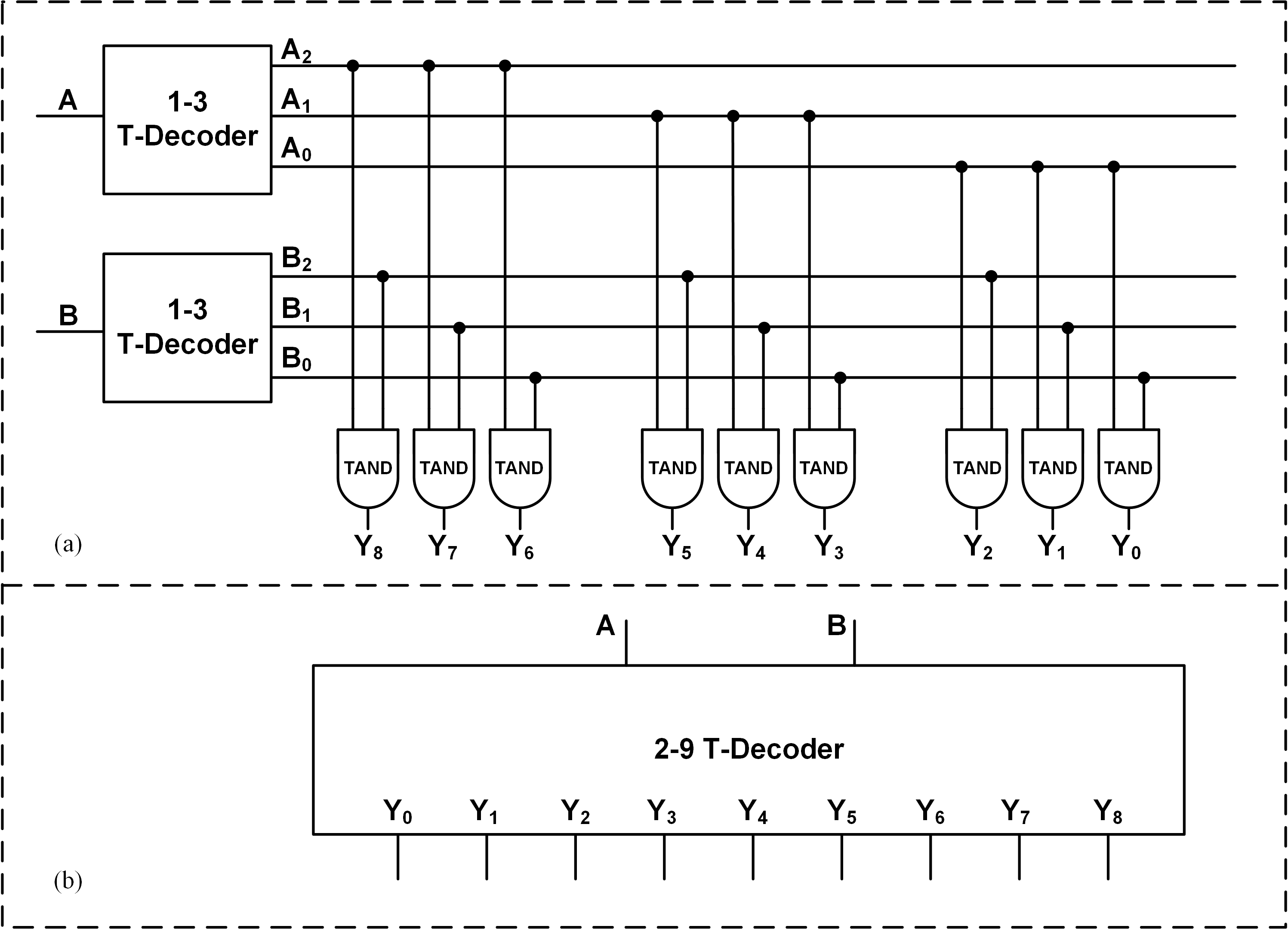}
\caption{Ternary 2-9 line decoder. (a) Gate level schematic using a pair of ternary 1-3 line decoders. TAND gates are constructed using the approach in Fig.~1(e) with source-followers used to buffer low-impedance nodes. (b) Block diagram.}
\label{fig4}
\end{figure*}

\begin{figure}[t!]
\centering
\includegraphics[scale=0.44]{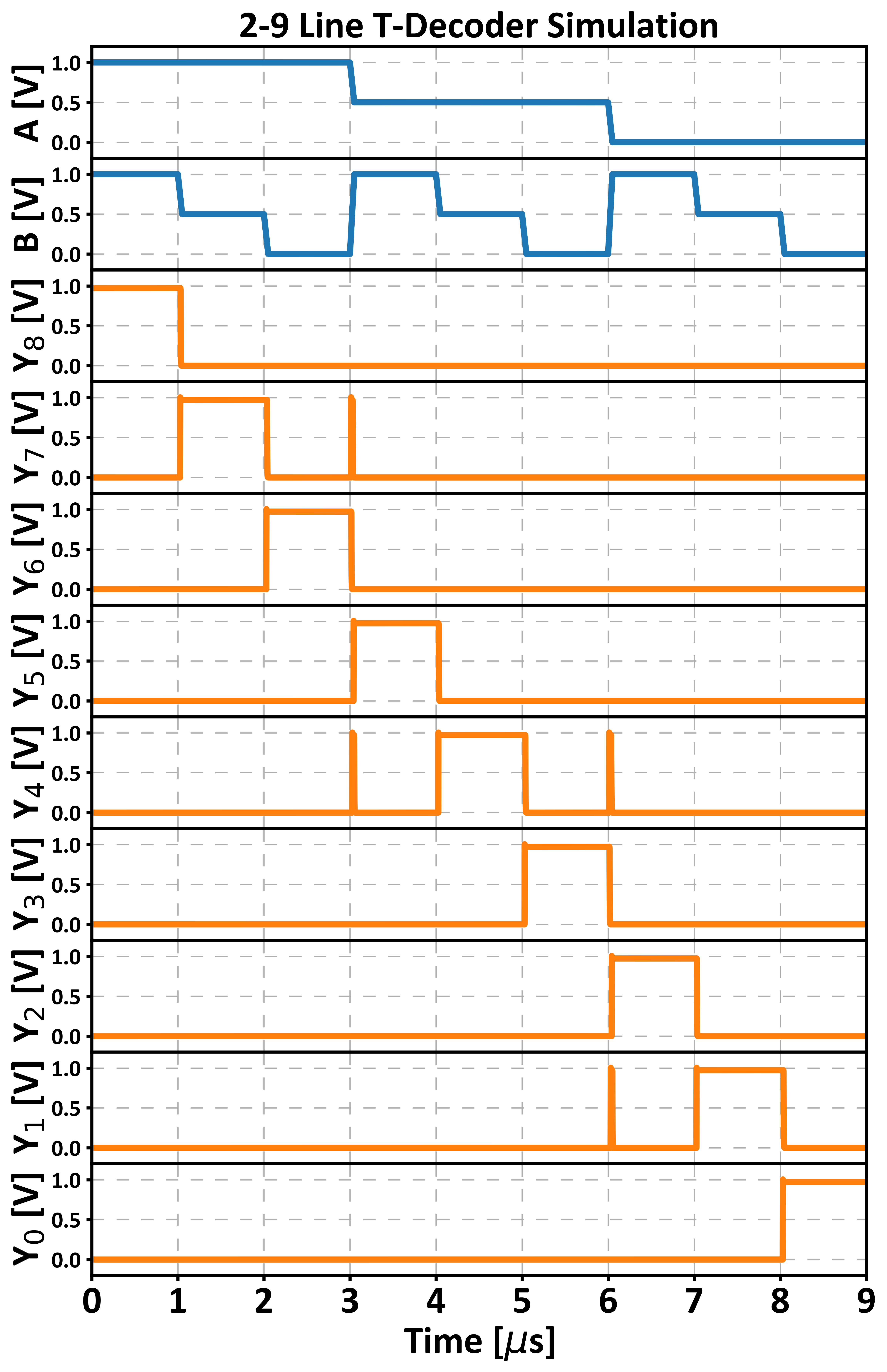}
\caption{SPICE simulation results of the memristor-CMOS ternary 2-9 line decoder from Fig.~4. Glitches occur due to competition hazards, which will marginally increase the hold time of each stage.}
\label{fig5}
\end{figure}

SPICE simulations were performed under the same conditions as the 1-3 line decoder, the results of which are provided in Fig.~5. The transients in the waveforms occur due to competition hazards arising from simultaneously changing inputs. The glitches in the output waveforms result in marginally longer hold times of each stage, such that the output signal has settled by the time it is captured.

\section{Display Decoder Design} 

Compared with a traditional digital 3-8 line decoder, the 2-9 line decoder uses fewer inputs to obtain more outputs owing to the ternary operation of the memristor-CMOS logic family. This improvement in data density is critical in high-resolution displays, and we use this circuit in a display decoder which can be used to drive a variety of display devices (e.g., LEDs, LCDs). Due to the ubiquity of seven-segment displays, we will design and implement a seven-segment display decoder with experimental results in the following section. The truth table of the ternary display decoder is provided in Table IV. 

\begin{table}[!t]\scriptsize
\centering\caption{Ternary Display Decoder Truth Table}
\begin{tabular}{
>{\columncolor[HTML]{FFFFFF}}l 
>{\columncolor[HTML]{FFFFFF}}l
>{\columncolor[HTML]{FFFFFF}}l
>{\columncolor[HTML]{FFFFFF}}l
>{\columncolor[HTML]{FFFFFF}}l 
>{\columncolor[HTML]{FFFFFF}}l
>{\columncolor[HTML]{FFFFFF}}l
>{\columncolor[HTML]{FFFFFF}}l
>{\columncolor[HTML]{FFFFFF}}l 
>{\columncolor[HTML]{FFFFFF}}l}
\hline\vspace{-6pt}
 &   \\ \hline
 
\multicolumn{1}{|c|}{\cellcolor[HTML]{FFFFFF}\textbf{$A$}} 
& \multicolumn{1}{c|}{\cellcolor[HTML]{FFFFFF}2} 
& \multicolumn{1}{c|}{\cellcolor[HTML]{FFFFFF}2}
& \multicolumn{1}{c|}{\cellcolor[HTML]{FFFFFF}2}
& \multicolumn{1}{c|}{\cellcolor[HTML]{FFFFFF}1}
& \multicolumn{1}{c|}{\cellcolor[HTML]{FFFFFF}1}
& \multicolumn{1}{c|}{\cellcolor[HTML]{FFFFFF}1}
& \multicolumn{1}{c|}{\cellcolor[HTML]{FFFFFF}0}
& \multicolumn{1}{c|}{\cellcolor[HTML]{FFFFFF}0}
& \multicolumn{1}{c|}{\cellcolor[HTML]{FFFFFF}0}\\ \hline

\multicolumn{1}{|c|}{\cellcolor[HTML]{FFFFFF}\textbf{$B$}} 
& \multicolumn{1}{c|}{\cellcolor[HTML]{FFFFFF}2} 
& \multicolumn{1}{c|}{\cellcolor[HTML]{FFFFFF}1}
& \multicolumn{1}{c|}{\cellcolor[HTML]{FFFFFF}0}
& \multicolumn{1}{c|}{\cellcolor[HTML]{FFFFFF}2}
& \multicolumn{1}{c|}{\cellcolor[HTML]{FFFFFF}1}
& \multicolumn{1}{c|}{\cellcolor[HTML]{FFFFFF}0}
& \multicolumn{1}{c|}{\cellcolor[HTML]{FFFFFF}2}
& \multicolumn{1}{c|}{\cellcolor[HTML]{FFFFFF}1}
& \multicolumn{1}{c|}{\cellcolor[HTML]{FFFFFF}0}\\ \hline

\multicolumn{1}{|c|}{\cellcolor[HTML]{FFFFFF}\textbf{$Y_a$}} 
& \multicolumn{1}{c|}{\cellcolor[HTML]{FFFFFF}0} 
& \multicolumn{1}{c|}{\cellcolor[HTML]{FFFFFF}0}
& \multicolumn{1}{c|}{\cellcolor[HTML]{FFFFFF}0}
& \multicolumn{1}{c|}{\cellcolor[HTML]{FFFFFF}0}
& \multicolumn{1}{c|}{\cellcolor[HTML]{FFFFFF}2}
& \multicolumn{1}{c|}{\cellcolor[HTML]{FFFFFF}0}
& \multicolumn{1}{c|}{\cellcolor[HTML]{FFFFFF}0}
& \multicolumn{1}{c|}{\cellcolor[HTML]{FFFFFF}2}
& \multicolumn{1}{c|}{\cellcolor[HTML]{FFFFFF}0}\\ \hline

\multicolumn{1}{|c|}{\cellcolor[HTML]{FFFFFF}\textbf{$Y_b$}} 
& \multicolumn{1}{c|}{\cellcolor[HTML]{FFFFFF}0} 
& \multicolumn{1}{c|}{\cellcolor[HTML]{FFFFFF}0}
& \multicolumn{1}{c|}{\cellcolor[HTML]{FFFFFF}2}
& \multicolumn{1}{c|}{\cellcolor[HTML]{FFFFFF}2}
& \multicolumn{1}{c|}{\cellcolor[HTML]{FFFFFF}0}
& \multicolumn{1}{c|}{\cellcolor[HTML]{FFFFFF}0}
& \multicolumn{1}{c|}{\cellcolor[HTML]{FFFFFF}0}
& \multicolumn{1}{c|}{\cellcolor[HTML]{FFFFFF}0}
& \multicolumn{1}{c|}{\cellcolor[HTML]{FFFFFF}0}\\ \hline

\multicolumn{1}{|c|}{\cellcolor[HTML]{FFFFFF}\textbf{$Y_c$}} 
& \multicolumn{1}{c|}{\cellcolor[HTML]{FFFFFF}0} 
& \multicolumn{1}{c|}{\cellcolor[HTML]{FFFFFF}0}
& \multicolumn{1}{c|}{\cellcolor[HTML]{FFFFFF}0}
& \multicolumn{1}{c|}{\cellcolor[HTML]{FFFFFF}0}
& \multicolumn{1}{c|}{\cellcolor[HTML]{FFFFFF}0}
& \multicolumn{1}{c|}{\cellcolor[HTML]{FFFFFF}0}
& \multicolumn{1}{c|}{\cellcolor[HTML]{FFFFFF}2}
& \multicolumn{1}{c|}{\cellcolor[HTML]{FFFFFF}0}
& \multicolumn{1}{c|}{\cellcolor[HTML]{FFFFFF}0}\\ \hline

\multicolumn{1}{|c|}{\cellcolor[HTML]{FFFFFF}\textbf{$Y_d$}} 
& \multicolumn{1}{c|}{\cellcolor[HTML]{FFFFFF}0} 
& \multicolumn{1}{c|}{\cellcolor[HTML]{FFFFFF}2}
& \multicolumn{1}{c|}{\cellcolor[HTML]{FFFFFF}0}
& \multicolumn{1}{c|}{\cellcolor[HTML]{FFFFFF}0}
& \multicolumn{1}{c|}{\cellcolor[HTML]{FFFFFF}2}
& \multicolumn{1}{c|}{\cellcolor[HTML]{FFFFFF}0}
& \multicolumn{1}{c|}{\cellcolor[HTML]{FFFFFF}0}
& \multicolumn{1}{c|}{\cellcolor[HTML]{FFFFFF}2}
& \multicolumn{1}{c|}{\cellcolor[HTML]{FFFFFF}0}\\ \hline

\multicolumn{1}{|c|}{\cellcolor[HTML]{FFFFFF}\textbf{$Y_e$}} 
& \multicolumn{1}{c|}{\cellcolor[HTML]{FFFFFF}0} 
& \multicolumn{1}{c|}{\cellcolor[HTML]{FFFFFF}2}
& \multicolumn{1}{c|}{\cellcolor[HTML]{FFFFFF}0}
& \multicolumn{1}{c|}{\cellcolor[HTML]{FFFFFF}2}
& \multicolumn{1}{c|}{\cellcolor[HTML]{FFFFFF}2}
& \multicolumn{1}{c|}{\cellcolor[HTML]{FFFFFF}2}
& \multicolumn{1}{c|}{\cellcolor[HTML]{FFFFFF}0}
& \multicolumn{1}{c|}{\cellcolor[HTML]{FFFFFF}2}
& \multicolumn{1}{c|}{\cellcolor[HTML]{FFFFFF}0}\\ \hline

\multicolumn{1}{|c|}{\cellcolor[HTML]{FFFFFF}\textbf{$Y_f$}} 
& \multicolumn{1}{c|}{\cellcolor[HTML]{FFFFFF}0} 
& \multicolumn{1}{c|}{\cellcolor[HTML]{FFFFFF}2}
& \multicolumn{1}{c|}{\cellcolor[HTML]{FFFFFF}0}
& \multicolumn{1}{c|}{\cellcolor[HTML]{FFFFFF}0}
& \multicolumn{1}{c|}{\cellcolor[HTML]{FFFFFF}0}
& \multicolumn{1}{c|}{\cellcolor[HTML]{FFFFFF}2}
& \multicolumn{1}{c|}{\cellcolor[HTML]{FFFFFF}2}
& \multicolumn{1}{c|}{\cellcolor[HTML]{FFFFFF}2}
& \multicolumn{1}{c|}{\cellcolor[HTML]{FFFFFF}0}\\ \hline

\multicolumn{1}{|c|}{\cellcolor[HTML]{FFFFFF}\textbf{$Y_g$}}
& \multicolumn{1}{c|}{\cellcolor[HTML]{FFFFFF}0} 
& \multicolumn{1}{c|}{\cellcolor[HTML]{FFFFFF}2}
& \multicolumn{1}{c|}{\cellcolor[HTML]{FFFFFF}0}
& \multicolumn{1}{c|}{\cellcolor[HTML]{FFFFFF}0}
& \multicolumn{1}{c|}{\cellcolor[HTML]{FFFFFF}0}
& \multicolumn{1}{c|}{\cellcolor[HTML]{FFFFFF}0}
& \multicolumn{1}{c|}{\cellcolor[HTML]{FFFFFF}0}
& \multicolumn{1}{c|}{\cellcolor[HTML]{FFFFFF}2}
& \multicolumn{1}{c|}{\cellcolor[HTML]{FFFFFF}2}\\ \hline

\multicolumn{1}{|c|}{\cellcolor[HTML]{FFFFFF}Display} 
& \multicolumn{1}{c|}{\cellcolor[HTML]{FFFFFF}\textifsym{8}} 
& \multicolumn{1}{c|}{\cellcolor[HTML]{FFFFFF}\textifsym{7}}
& \multicolumn{1}{c|}{\cellcolor[HTML]{FFFFFF}\textifsym{6}}
& \multicolumn{1}{c|}{\cellcolor[HTML]{FFFFFF}\textifsym{5}}
& \multicolumn{1}{c|}{\cellcolor[HTML]{FFFFFF}\textifsym{4}}
& \multicolumn{1}{c|}{\cellcolor[HTML]{FFFFFF}\textifsym{3}}
& \multicolumn{1}{c|}{\cellcolor[HTML]{FFFFFF}\textifsym{2}}
& \multicolumn{1}{c|}{\cellcolor[HTML]{FFFFFF}\textifsym{1}}
& \multicolumn{1}{c|}{\cellcolor[HTML]{FFFFFF}\textifsym{ 0}}\\ \hline

\hline \vspace{-6pt}
  &  \\ \hline
\end{tabular}%
\end{table}

According to the truth table of the ternary display decoder, the following can be concluded:

\begin{align*}
     Y_a &= Y_1 + Y_4
     & Y_b = Y_5 + Y_6 \\
     Y_c &= Y_2
     &Y_d = Y_1 + Y_4+Y_7\\
\end{align*}  \vspace{-1.85cm}

\begin{align*}
     Y_e &=Y_1+Y_3+Y_4+Y_5+Y_7\\
     Y_f &= Y_1+Y_2+Y_3+Y_7\\
     Y_g &= Y_0+Y_1+Y_7,
\end{align*}

\noindent where $Y_{a-g}$ are the the outputs of the display decoder, and $Y_1-Y_8$ are the outputs of the 2-9 line decoder. Therefore, the display decoder can be realized by routing the 2-9 line decoder through TOR gates as shown in Fig.~6. As before, $A$ and $B$ are the two inputs fed into the 2-9 ternary decoder from Fig.~4, and the outputs $Y_{a-g}$ are connected to the common anode of the seven-segment display, generating the displays shown in the bottom row of Table IV. SPICE simulation results of the full 2-9 ternary decoder, including source-follower buffers to cascade the various stages together, are shown in Fig.~7. 

\begin{figure}[t!]
\centering
\includegraphics[scale=0.71]{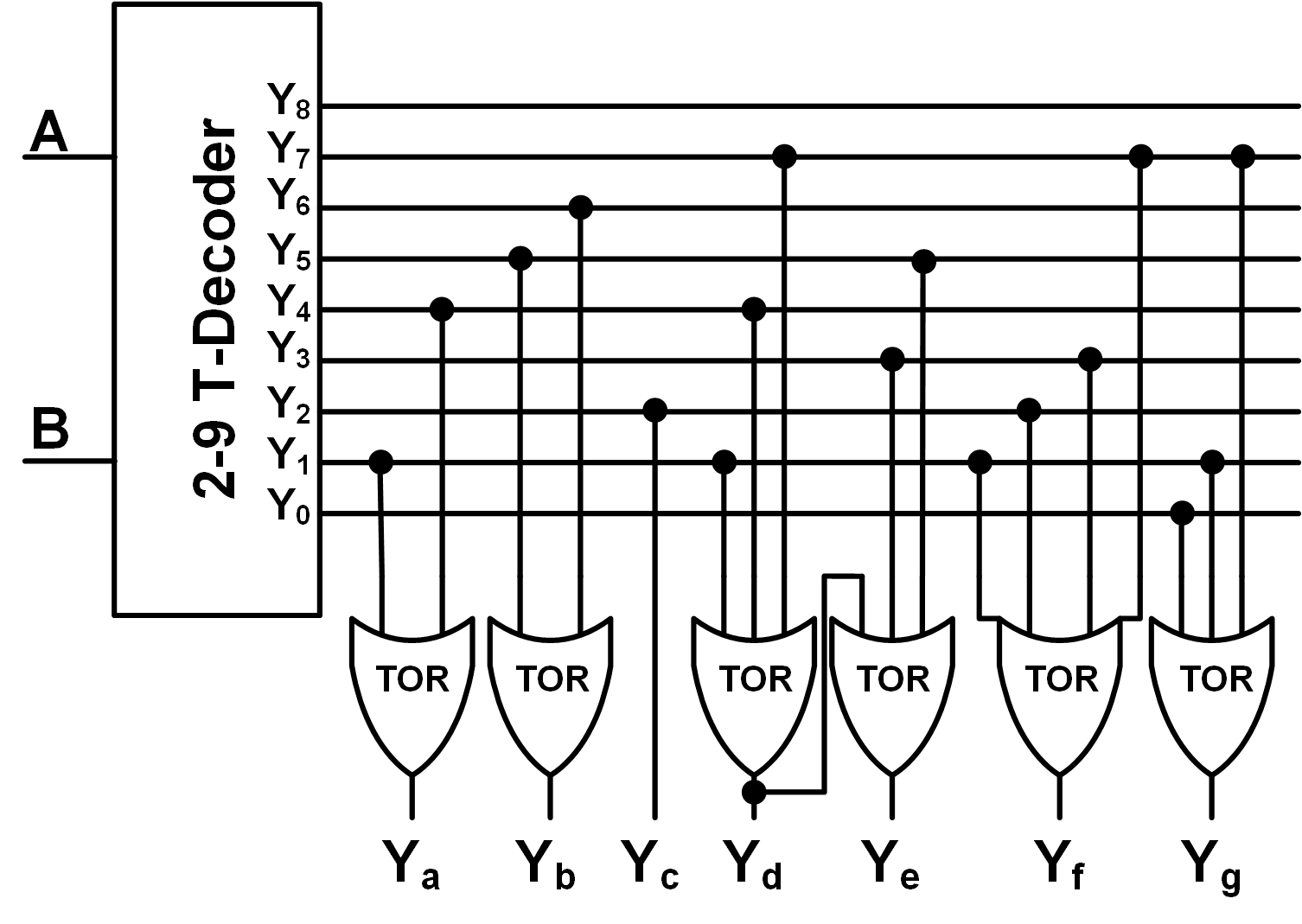}
\caption{Schematic of memristor-CMOS ternary display decoder consisting of a 2-9 ternary line decoder (Fig.~4), and TOR gates (Fig.~1(f)).}
\label{fig4}
\end{figure}

\begin{figure}[t!]
\centering
\includegraphics[scale=0.43]{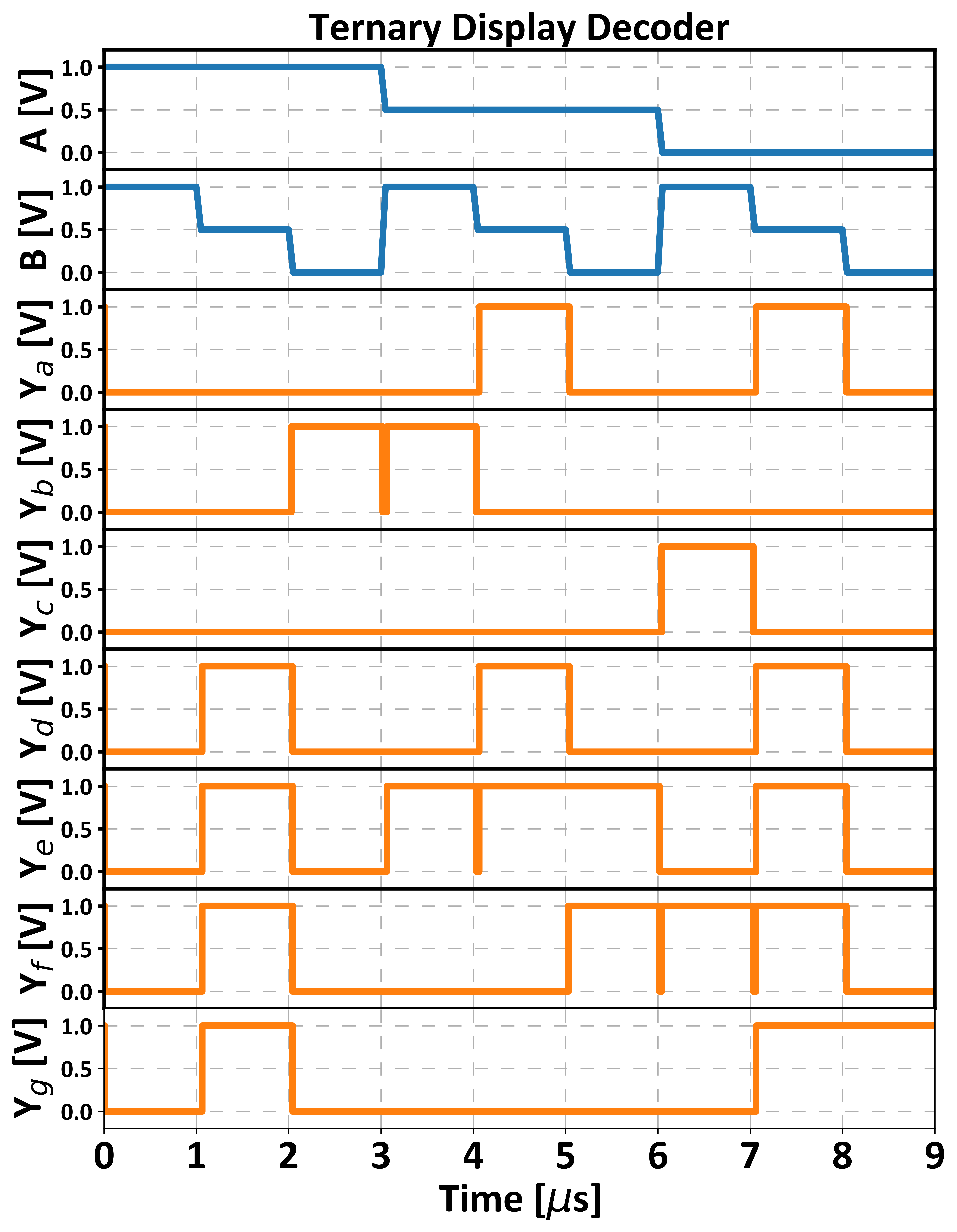}
\caption{SPICE simulation results of the memristor-CMOS ternary display decoder from Fig.~6. These signals are passed to the common anode of a seven-segment display in the next section.}
\label{fig4}
\end{figure}

\begin{figure}[!t]
\centering
\subfloat[]
{
	\includegraphics[scale=0.75]{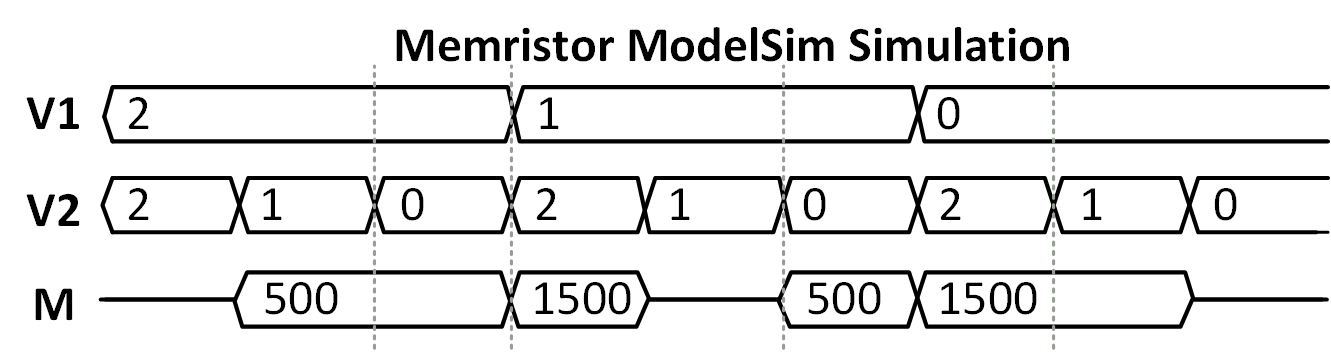}
	\label{fig3a}
}\\
\subfloat[]
{
	\includegraphics[scale=0.75]{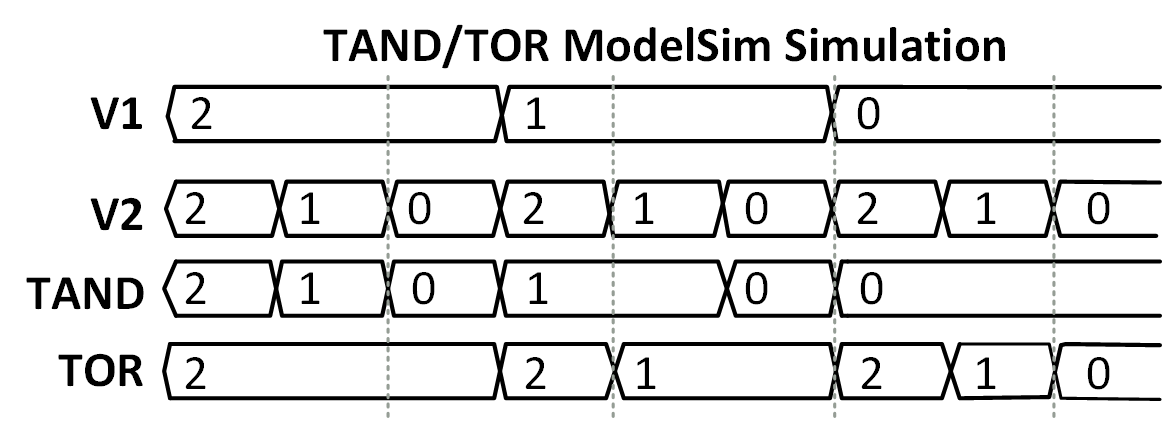}
	\label{fig3b}
}\\
\subfloat[]
{
	\includegraphics[scale=0.75]{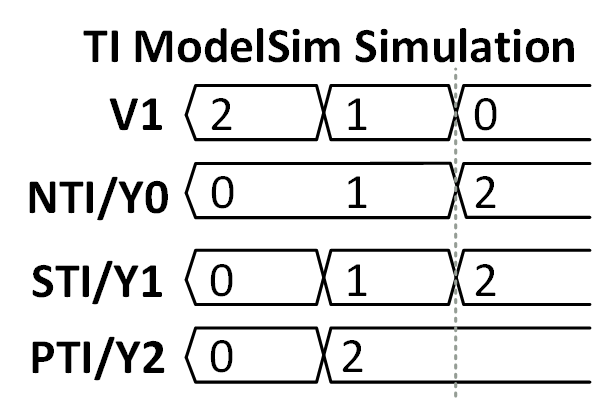}
	\label{fig3b}
}
\caption{ModelSim Simulations: Logic Primitives (a) Memristor Model. (b) TAND and TOR Gates. (c) NTI, STI and PTI Gates.}
\label{fig8}
\end{figure}

\section{FPGA Simulation and Experimental Results} 
\subsection{ModelSim Simulation}
The lack of a native memristor array on FPGA meant that we had to first develop a memristor model in Quartus II. The logic gates used only require the memristors to be switched between two states, and so a digital memristor model is used. We take an approach analogous to current EDA tools that integrate RRAM and MRAM macros \cite{chiu201940nm, gallagher2019recent, gallagher201922nm}. 

The anode (positive terminal) of the memristor is represented with $V1$ and the cathode (negative terminal) with $V2$. The memristance is treated as an output. When the voltage is forward biased ($V1 > V2$), the output is set to ``500'' as a binary representation of R$_{\rm on}$. When the voltage is reverse-biased, the output is set to ``1500'' as a binary representation of $R_{\rm off}$. A low resistance ratio is used here for short bit-widths, though in practice should be larger to maintain good noise immunity as in the analog simulations shown in Figs.~3, 5, and 7.

The corresponding ModelSim simulation results of the memristor is shown in Fig.~8(a). The memristance of the memristor is displayed in decimal. In a similar way, TAND, TOR and TI gates are programmed on Quartus II. The TAND and TOR simulation results are provided in Fig.~8(b), and the ternary inverters (NTI, STI and PTI) in Fig.~8(c). Ternary outputs are represented in the digital environment and development board by using an additional bit: (0, 1, 2) = (00, 01 10).

The ModelSim simulation results of the three ternary decoders are shown in Fig.~9. Note that the display decoder outputs a high level of `2' (10) and `0' (00) in the ternary domain. These two bits are OR'd to convert them into a value interpretable by the seven-segment display. Having demonstrated the correct operation of all combinational logic circuits in ModelSim, the HDL netlists are ready for FPGA synthesis.

\begin{figure}[!t]
\centering
\subfloat[]
{
	\includegraphics[scale=0.75]{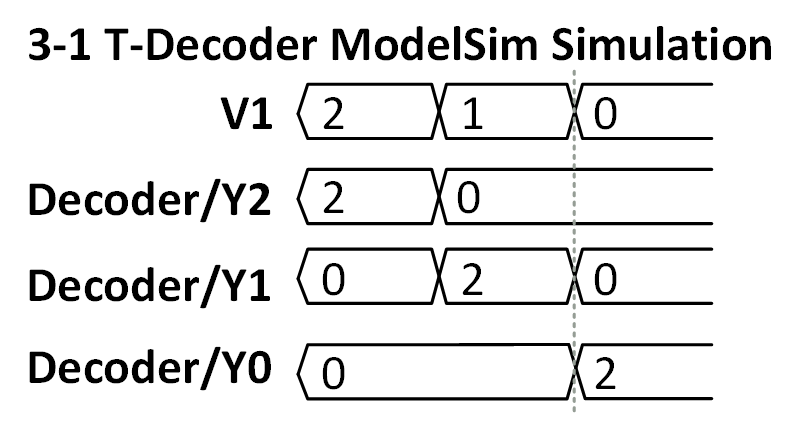}
	\label{fig3a}
}\\
\subfloat[]
{
	\includegraphics[scale=0.75]{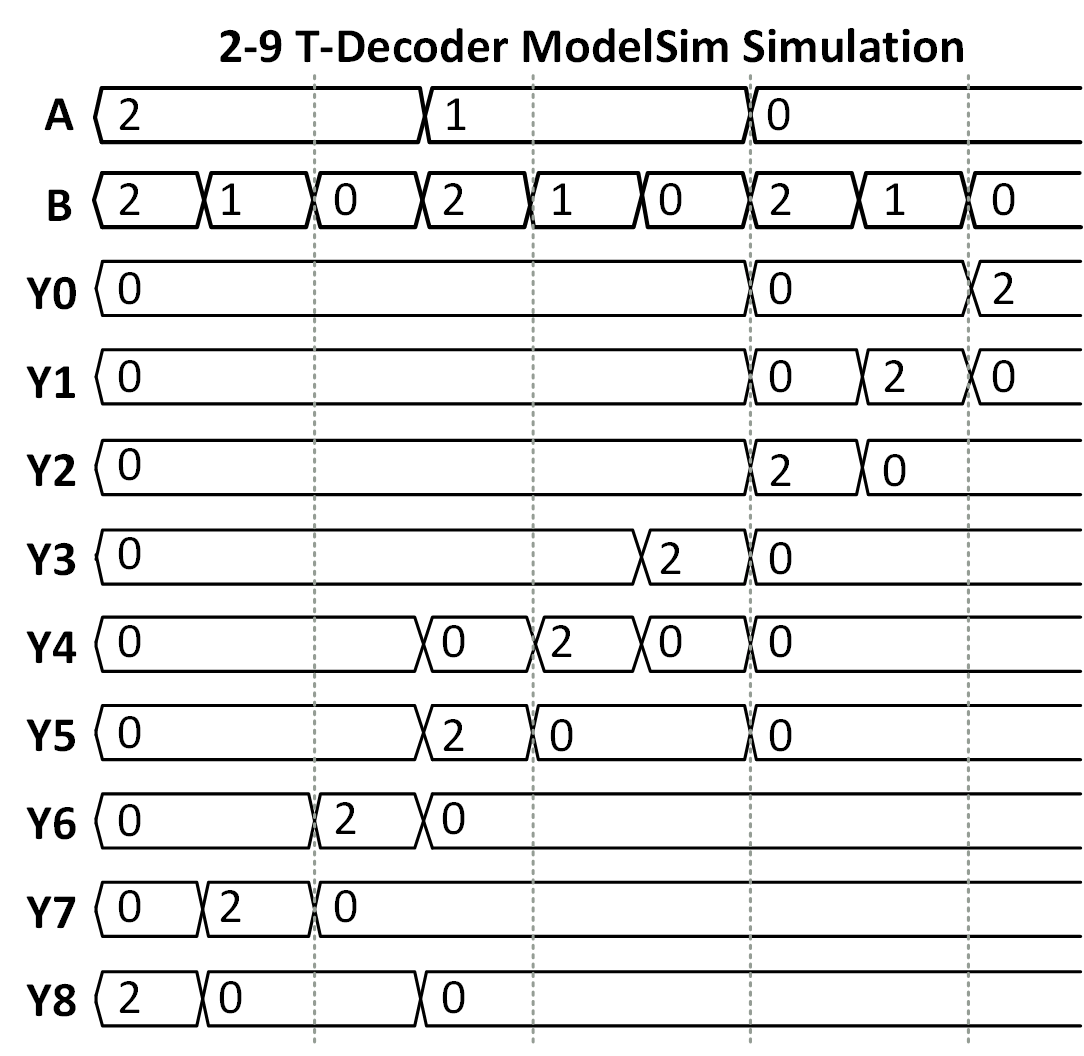}
	\label{fig3b}
}\\
\subfloat[]
{
	\includegraphics[scale=0.75]{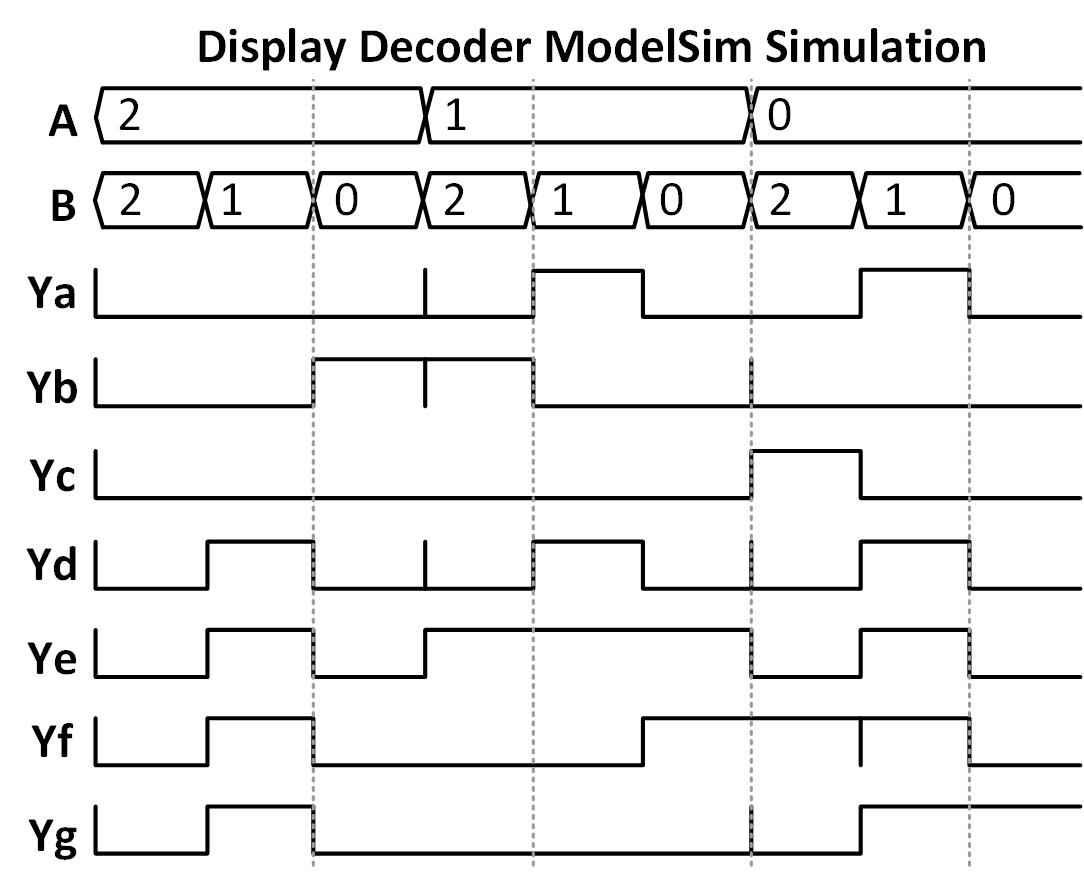}
	\label{fig3b}
}
\caption{ModelSim Simulations: Ternary Decoders (a) 1-3 T-Decoder. (b) 2-9 T-Decoder. (c) Ternary Display Decoder.}
\label{fig8}
\end{figure}

\subsection{FPGA Experimental Results}
The HDL netlist is synthesized on an Altera Cyclone IV EP4CE6E22 development board shown in Fig.~10(a). The input signals are controlled by on-board DIP switches labeled (i), and the seven-segment LED display labelled (ii) changes accordingly. All combinations of inputs $A$ and $B$ with the corresponding display response are shown in Fig.~10(b), with the timing diagram in Fig.~9(c), thus verifying the physical operation of the combinational memristor-CMOS ternary logic scheme.

\begin{figure*}[t!]
\centering
\includegraphics[scale=0.75]{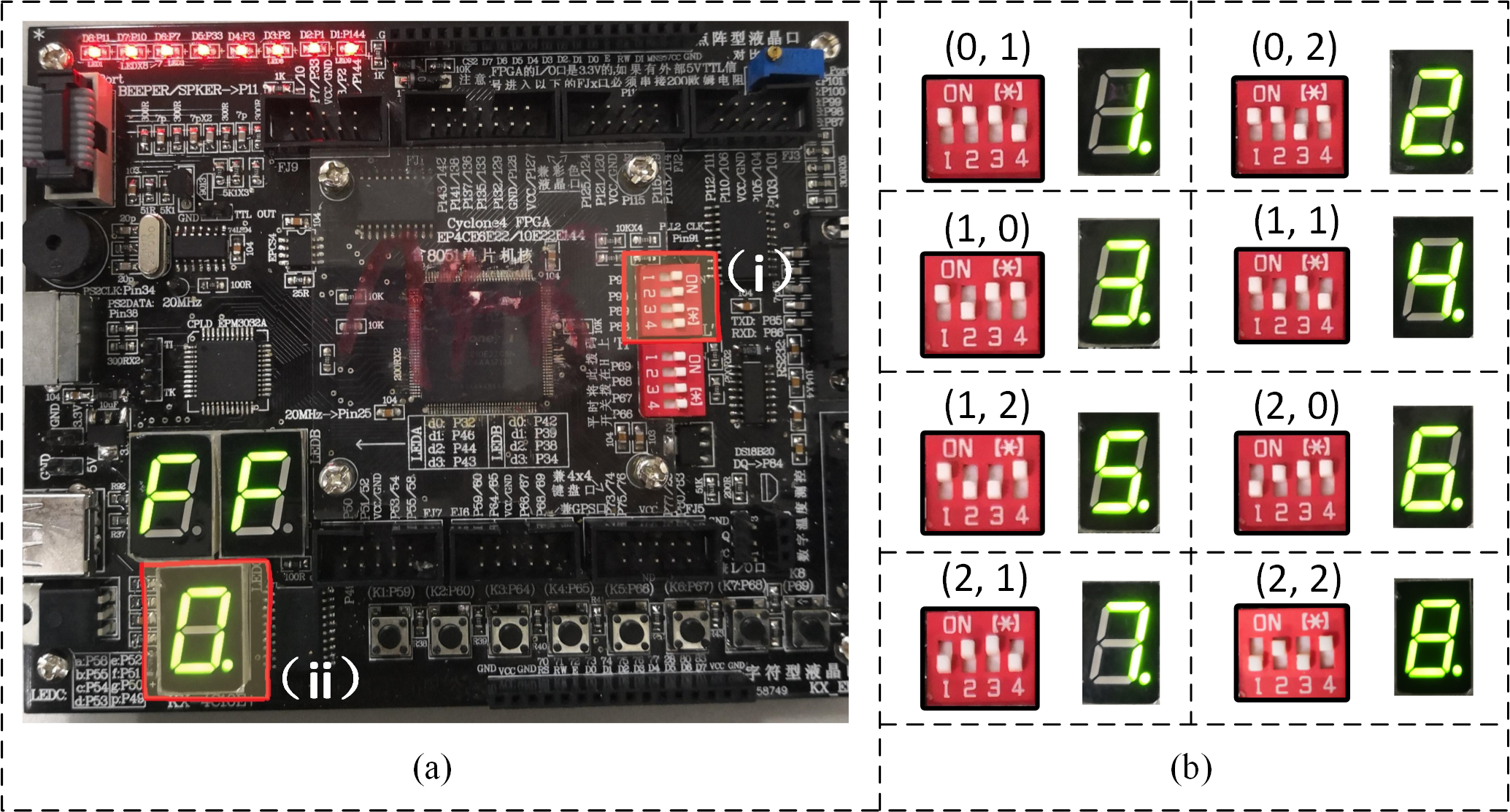}
\caption{Altera Cyclone IV EP4CE6E22 FPGA development board running the memristor-CMOS ternary display decoder. (a) Memristors are synthesized as conditional binary switches. Inputs are manually altered in (i) using DIP switches, changing the status of the seven-segment display in (ii). Pictured on the left is for the case of an input signal of (0, 0). (b) Each input configuration with the associated output.}
\label{fig4}
\end{figure*}

\subsection{Discussion and Comparison}


The memristor-CMOS display driver was calculated to consume 2mW of I/O power, and 60mW of average static power. The parameters generated by the PowerPlay Early Power Estimator are provided in Table V. This was compared to a baseline BCD-seven segment decoder implemented in standard Boolean logic, where an equivalent analysis resulted in I/O power of 14~mW, and equivalent static power of 60~mW. Our memristor-CMOS approach reduces I/O power by a factor of six times, and the total FPGA power by 18.75\%. This is most likely a result of ternary-domain processing reducing the total required I/O.

The FPGA device utilization summary shows that our approach requires 154 look-up tables (LUTs), 154 flip-flops (FFs), 11 registers, and 13 total pins. The digital baseline uses 26 LUTs, 26 FFs, 12 registers, and 17 total pins. Although our approach requires many more LUTs and FFs, most of this overhead is due to the absence of an integrated RRAM array. Regardless of the additional resources, our approach requires four less pins, which is where most of the power overhead comes from. This saving implies that there are cases where on-board RRAM can significantly improve resource management on development boards. Overall, the data density advantages of memristor-CMOS logic remain even when the memristor is emulated. 

To measure the maximum operating speed, the display decoder was treated as a single stage with registers before and after in the design. The maximum synthesizable frequency competition hazards corrupted the output was 293.8~MHz based on the timing report. The maximum synthesizable frequency obtained is 577.7~MHz, which is a factor of 1.98 times faster than our approach. Realistically, this approach is estimated to slow down by approximately an order of magnitude when the synthesized memristor model is interchanged with an RRAM device \cite{wang2020high}. This trade-off was to be expected, which is precisely why it is applied as a display decoder where high speed operation is an ancillary metric when considering density, resolution and fill-factor.

\begin{table}[!t]\scriptsize
\centering\caption{FPGA Power Estimation$^1$}
\begin{tabular}{
>{\columncolor[HTML]{FFFFFF}}l 
>{\columncolor[HTML]{FFFFFF}}l}
\hline\vspace{-6pt}

 &   \\ \hline
\multicolumn{1}{|c|}{\cellcolor[HTML]{FFFFFF}\textbf{Parameter}} & 
\multicolumn{1}{c|}{\cellcolor[HTML]{FFFFFF}\textbf{Value}} \\ \hline

\multicolumn{1}{|c|}{\cellcolor[HTML]{FFFFFF}Family} & 
\multicolumn{1}{c|}{\cellcolor[HTML]{FFFFFF} Cyclone IV E} \\ 
 
\multicolumn{1}{|c|}{\cellcolor[HTML]{FFFFFF}Device} & 
\multicolumn{1}{c|}{\cellcolor[HTML]{FFFFFF} EP4CE6}\\ 

\multicolumn{1}{|c|}{\cellcolor[HTML]{FFFFFF}Package} & 
\multicolumn{1}{c|}{\cellcolor[HTML]{FFFFFF} E22} \\ 

\multicolumn{1}{|c|}{\cellcolor[HTML]{FFFFFF}Temperature Grade} & 
\multicolumn{1}{c|}{\cellcolor[HTML]{FFFFFF} Commercial} \\ 

\multicolumn{1}{|c|}{\cellcolor[HTML]{FFFFFF}Power Characteristics} & 
\multicolumn{1}{c|}{\cellcolor[HTML]{FFFFFF} Typical}\\ 

\multicolumn{1}{|c|}{\cellcolor[HTML]{FFFFFF}V$_{\rm SS}$} & 
\multicolumn{1}{c|}{\cellcolor[HTML]{FFFFFF} 1.20 V}\\

\multicolumn{1}{|c|}{\cellcolor[HTML]{FFFFFF}Ambient Temp.} & 
\multicolumn{1}{c|}{\cellcolor[HTML]{FFFFFF} 298 K}\\

\multicolumn{1}{|c|}{\cellcolor[HTML]{FFFFFF}V$_{\rm SS}$} & 
\multicolumn{1}{c|}{\cellcolor[HTML]{FFFFFF} 1.20 V}\\ \hline

\multicolumn{1}{|c|}{\cellcolor[HTML]{FFFFFF}\textbf{Thermal Power}} & 
\multicolumn{1}{c|}{\cellcolor[HTML]{FFFFFF}}\\ \hline

\multicolumn{1}{|c|}{\cellcolor[HTML]{FFFFFF}I/O} & 
\multicolumn{1}{c|}{\cellcolor[HTML]{FFFFFF} 2 mW}\\

\multicolumn{1}{|c|}{\cellcolor[HTML]{FFFFFF}Static} & 
\multicolumn{1}{c|}{\cellcolor[HTML]{FFFFFF} 60mW}\\

\multicolumn{1}{|c|}{\cellcolor[HTML]{FFFFFF}Total FPGA} & 
\multicolumn{1}{c|}{\cellcolor[HTML]{FFFFFF} 62mW}\\
 
\hline \vspace{-6pt}
  &  \\ \hline 
  \multicolumn{2}{c}{\scriptsize{$^1$Extracted with Altera PowerPlay Early Power Estimator}}
\end{tabular}%
\end{table}

Beyond FPGA synthesis, memristor integration in the BEOL of microLED displays is promising not only due to the high density of hybrid RRAM-CMOS technologies, but also the process compatibility of LED technologies with emerging approaches to fabricating RRAM. Our prior experimental demonstration of ternary memristor-CMOS logic in \cite{wang2020high} uses indium-tin-oxide (ITO) as the switching layer, and indium is commonly used in the manufacturing process of both CMOS and gallium-nitride (GaN)-based microLED arrays \cite{mckendry2009individually}. The work in \cite{chen2019fabrication} uses RF sputtering of a 70-nm thick layer of ITO on the wafer, used as the current spreading layer to ensure an even distribution of charge injection across the active layer on the substrate. Synonymously, our indium-based RRAM device required RF-sputtering 10-nm thin film layer of ITO, resulting in a planar area occupation of 0.6$\mu$m $\times$0.6$\mu$m which is highly compatible with LED arrays of submicron pitch.

\section{Conclusion}
Novel ternary memristor-CMOS decoders are proposed, with their analog characteristics simulated in SPICE, and synthesized on an FPGA development board. In this design, the number of memristors used is relatively large, and so FPGA experiments are used instead of device-level characterization. We demonstrate a reduction of total power in the synthesized results as a result of processing in the ternary domain, which is compatible with memristor-CMOS logic. This can push towards the practical use of emerging logic families in applications where speed may be safely foregone in favor of data density.



\bibliographystyle{IEEEtran}
\bibliography{references}

\end{document}